\documentclass[lettersize,journal]{IEEEtran}
\usepackage{amsmath,amsfonts}
\usepackage{amssymb}
\usepackage{algorithmic}
\usepackage{algorithm}
\usepackage{array}
\usepackage[caption=false,font=normalsize,labelfont=sf,textfont=sf]{subfig}
\usepackage{textcomp}
\usepackage{stfloats}
\usepackage{hyperref}
\usepackage{url}
\usepackage{verbatim}
\usepackage{graphicx}
\usepackage{cite}
\usepackage{tabularray}
\usepackage[table]{xcolor}
\newtheorem{theorem}{Theorem}

\begin{document}

\title{RFCheck: Synthetic RF Sensing Data Can Fail Measurement Consistency}

\author{Di Zhang,~\IEEEmembership{Student Member,~IEEE,}
        Yuanhao Cui,~\IEEEmembership{Member,~IEEE,} Tony Xiao Han,~\IEEEmembership{Senior Member,~IEEE,}
        \\and Xiaojun Jing,~\IEEEmembership{Member,~IEEE}
\thanks{Di Zhang, Yuanhao Cui, and Xiaojun Jing are with the School of Information and Communication Engineering, Beijing University of Posts and Telecommunications, Beijing 100876, China (e-mails: amandazhang@bupt.edu.cn, yuanhao.cui@bupt.edu.cn, jxiaojun@bupt.edu.cn).}
\thanks{Tony Xiao Han is with Huawei Technologies Co., Ltd., Shenzhen, Guangdong 518129, China (e-mail: tony.hanxiao@huawei.com).} }

\markboth{IEEE Transactions on Mobile Computing,~Vol.~XX, No.~X, 2026}%
{Zhang \MakeLowercase{\textit{et al.}}: RFCheck for Synthetic RF Sensing Data}

\maketitle

\begin{abstract}
Synthetic radio-frequency (RF) sensing data are widely used to augment wireless sensing tasks, yet their measurement behavior is often not directly evaluated under the acquisition conditions in which they are used. This paper identifies a measurement-consistency failure mode in synthetic RF sensing data. A synthetic RF sample may appear useful to a classifier and pass common task-facing checks while deviating from the measurement behavior of real samples collected and processed by the same sensing pipeline. Such deviations can introduce synthetic shortcuts and bias downstream model selection.

We study this failure under a matched acquisition setting, where the reference is not an ideal physical signal but held-out real data produced by the same acquisition and preprocessing pipeline. To make the failure observable, we instantiate RFCheck as a calibrated measurement audit. It calibrates representation-specific measurement tests on held-out real samples and flags synthetic samples whose responses exceed the calibrated real-data range. We then use the audit to screen candidates. Retention tests whether low-risk candidates behave differently from high-risk ones. Repair and correction further test whether the detected violations can be reduced for proposal pools that already contain task-relevant sensing structure.

Channel state information (CSI) is the primary validation setting. The CSI audit checks whether synthetic samples preserve delay-domain and local frequency-domain structure induced by real measurements. Wi-Fi CSI experiments show that aggregate summaries and label-based screening can miss measurement failures detected by the calibrated audit. Under the same label acceptance rule, low-risk and high-risk synthetic candidates lead to different task behavior. A repair reference reduces the flagged ratio to $10.83\%$ while preserving mean task performance. In a held-out proposal study, correction followed by calibrated selection obtains a class-balanced set with no flagged sample under the fixed training budget. These intervention results show that measurement violations can be reduced when the proposal pool already contains task-relevant structure.

We further instantiate the same calibration principle on frequency-modulated continuous-wave (FMCW) millimeter-wave radar gesture sensing. These results support the central claim of this paper. Synthetic RF sensing data can contain measurement failures that common task checks do not expose. Such failures should be diagnosed before augmentation and reduced when the proposal pool contains usable sensing structure.

\end{abstract}


\section{Introduction}
\label{sec:intro}

Integrated sensing and communication (ISAC) is expected to support both wireless connectivity and environment sensing in future wireless systems \cite{9737357,zhang2026integrated,wang2015understanding}. As wireless sensing becomes increasingly data-driven, synthetic radio-frequency (RF) sensing data are receiving growing attention for model training and data augmentation \cite{10437154,wen2025generative,van2024generative,hou2024rfboost,chi2024rf,letafati2023diffusion}. This demand is practical. Large real sensing datasets are expensive to collect, and many deployment conditions are hard to sample densely.

Synthetic sensing data do not always help. The failure studied in this paper is not simply a wrong label or a large summary distance. It is measurement inconsistency. A synthetic RF sample may be accepted by a task model and may look close to real data under selected summaries, yet still violate the structure imposed by the real sensing pipeline.

This failure takes a specific form in RF sensing. A channel-state-information (CSI) sample or radar tensor is not just an image-like array. It is produced by a finite-bandwidth measurement process and then transformed by benchmark-specific preprocessing. The occupied spectrum and receiver response leave structure in the final representation. If a synthetic sample ignores this structure, a task model may learn proposal artifacts instead of sensing behavior. The apparent gain from augmentation can therefore be misleading.

This issue is especially important for CSI, one of the most widely used RF sensing representations \cite{tan2022commodity,chen2023cross,wen2024survey,zhu2025csi}. CSI is available on commodity wireless platforms and is widely used in gesture and activity recognition. Recent RF sensing benchmarks have also moved toward larger action sets and safety-oriented tasks \cite{wang2024xrf55,lan2024bullydetect}. Accordingly, this paper uses CSI as the primary validation setting. The central question is whether synthetic CSI remains consistent with the held-out real measurement baseline before it is used for augmentation.

Common validation routes do not fully expose this failure mode. Aggregate distances compare distributions after summary features are chosen, so they may miss localized measurement violations. Label-consistency checks only test whether a classifier accepts the intended class. Downstream accuracy is even less direct because the model is trained before the failure is observed. These checks remain useful, but they do not answer whether a synthetic RF sample is consistent with how real measurements are produced.

\begin{figure*}[!htp]
\centering
\includegraphics[width=0.95\textwidth]{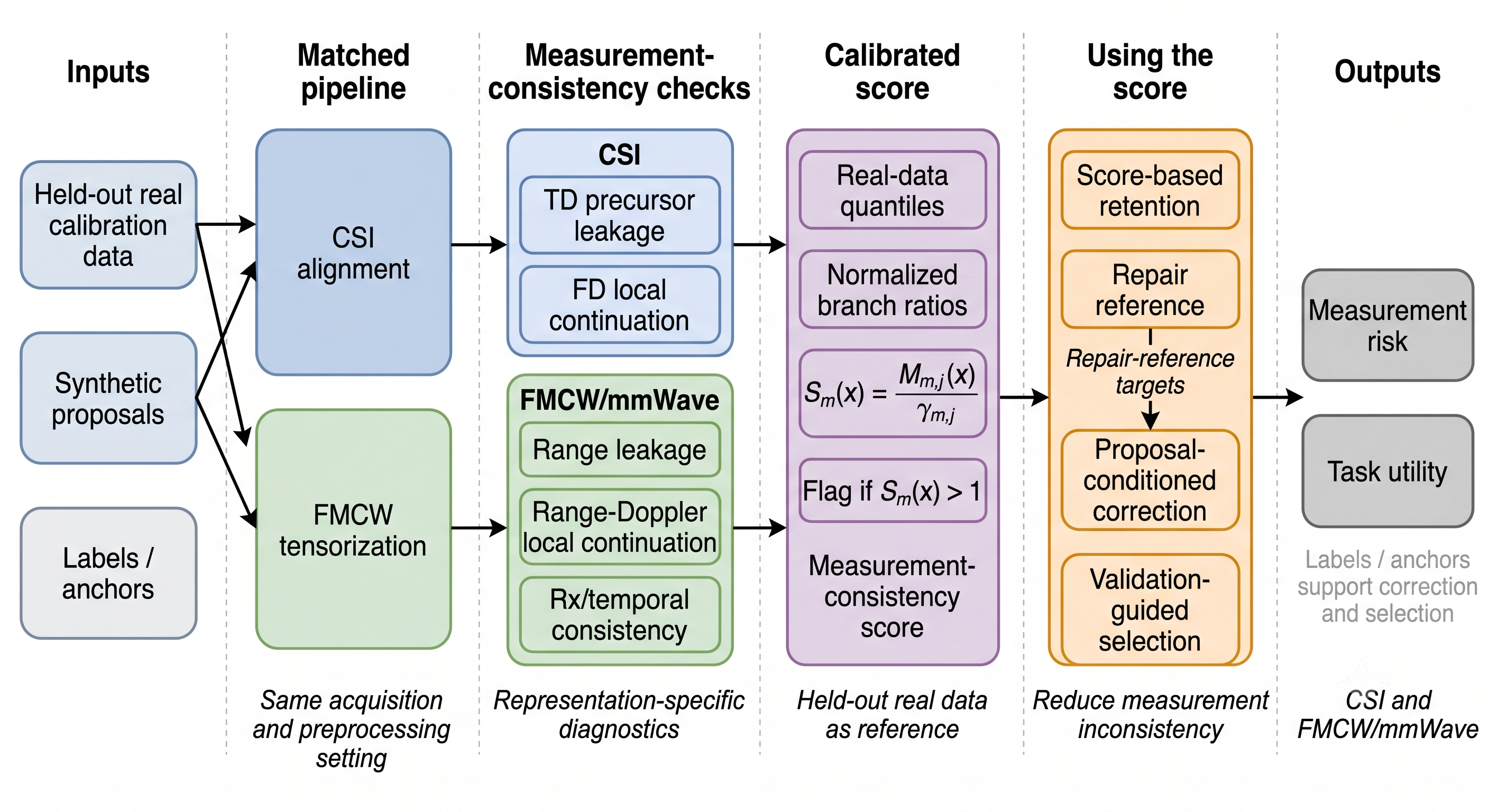}
\caption{From synthetic RF failure to measurement-consistent correction. Matched real measurements define the reference, RFCheck detects measurement violations, and ranking or correction reduces the failure for suitable proposal pools.}
\label{fig:rfcheck_overview}
\end{figure*}

We study this problem under a matched acquisition setting. The reference is not an ideal physical channel or an ideal radar response. The reference is the held-out real data produced by the same sensing pipeline. A synthetic sample is treated as risky when its measurement response exceeds the calibrated range of this real reference.

We instantiate this matched measurement audit as RFCheck. It computes measurement tests, calibrates them on held-out real samples, and assigns each synthetic candidate a measurement-consistency score. The audit is used to diagnose candidates, and repair or correction is then used to test whether the detected failure can be reduced. CSI is used as the primary validation setting, while frequency-modulated continuous-wave (FMCW) millimeter-wave gesture sensing tests whether the same principle can be instantiated with different measurement tests. Figure~\ref{fig:rfcheck_overview} summarizes the evaluation flow.

The main contributions of this paper are summarized below.

\begin{itemize}

\item We identify measurement inconsistency as a failure mode of synthetic RF sensing data. 
A synthetic RF sample may pass common task checks, including selected summary statistics or label-consistency tests, while still deviating from the measurement behavior of held-out real samples produced by the same acquisition and preprocessing pipeline.

\item We empirically characterize how this failure escapes common validation routes. 
Controlled CSI perturbations and source-audit experiments show that aggregate summary distances do not provide sample-level measurement localization, and that label-consistent synthetic candidates can still carry measurement risk that affects downstream behavior under a fixed class budget.

\item We make this failure measurable through a matched measurement audit. RFCheck calibrates representation-specific measurement tests on held-out real data and exposes sample-level exceedances relative to the empirical measurement range. For CSI, the audit checks whether synthetic samples preserve delay-domain and local frequency-domain structure induced by real measurements. 
For FMCW sensing, the same calibration logic is applied to range-Doppler measurement structure.

\item We use retention, repair, and correction as interventions to test when the identified failure can be reduced. 
Under the same label acceptance rule, calibrated feedback separates low-risk candidates from high-risk ones. The repair reference and correction study then show that violations can be reduced when the original candidates already preserve task-relevant sensing structure. Synthetic sources with limited task-relevant sensing structure remain limited, indicating that calibrated correction is proposal-dependent: it can reduce measurement-side violations, but it cannot recover task-relevant sensing structure that is absent from the original candidates.

\end{itemize}

The scope of this paper is measurement consistency for synthetic RF sensing samples under matched acquisition and preprocessing settings. It does not replace task evaluation or hardware-level RF validation. Instead, it complements them by checking whether synthetic samples remain within the empirical measurement range of held-out real data before they are used for augmentation.

\section{Related Work}
\label{sec:related_work}

\textit{Synthetic RF sensing data.}
Synthetic RF and CSI data have been studied for augmentation and sim-to-real transfer in wireless sensing \cite{10437154,wen2025generative,van2024generative,hou2024rfboost,chi2024rf}. Recent RF generative models, including RF-Diffusion \cite{chi2024rf}, focus on realistic radio signals and task performance. This paper addresses a different question: whether synthetic sensing samples remain within the measurement behavior of held-out real data. Simulation-based channel datasets and physics-informed channel generators provide another important line of work \cite{alkhateeb2019deepmimo,bock2025physics}, but they target channel modeling rather than sample-level checking for measured sensing representations.

\textit{Similarity and quality filtering.}
Distance measures such as maximum mean discrepancy (MMD) are commonly used to compare real and synthetic data \cite{gretton2012kernel}. They are useful for aggregate matching, but they do not directly test whether each sample respects measurement-side structure under the matched acquisition and preprocessing setting. Quality-guided filtering provides a related idea for wireless synthetic data \cite{gong2025data}, but RFCheck differs by using measurement tests tied to the sensing representation and calibrated with real data from the same benchmark.

\textit{Measurement-side plausibility under matched acquisition settings.}
CSI samples are shaped by the measurement chain before they become sensing tensors \cite{tan2022commodity,wen2024survey,zhu2025csi}. This chain leaves structure in the occupied spectrum and phase behavior. FMCW tensors also carry measurement structure because range-Doppler resolution constrains how motion appears in the final representation. Prior millimeter-wave gesture studies have shown that range-Doppler and temporal motion structure are central to robust gesture recognition \cite{liu2021m,hayashi2021radarnet,grobelny2022mm}. For this reason, the relevant reference is not an idealized physical signal, but the empirical baseline induced by the same acquisition and preprocessing setting. RFCheck follows this view by checking whether synthetic samples deviate from held-out real measurements beyond calibrated ranges.

\textit{Scope of represented modalities.}
The present paper uses CSI as the primary sensing representation and FMCW gesture sensing as a second RF setting. Other RF modalities are left to future work.

\section{Measurement Consistency Under Matched RF Measurements}
\label{sec:problem_framework}
This section describes how the identified measurement-consistency failure is diagnosed and reduced. The goal is to operationalize a matched measurement audit for synthetic sensing samples. Held-out real measurements collected under the same acquisition and preprocessing setting define the empirical reference. Synthetic samples are treated as risky when their representation-specific measurement responses exceed the calibrated real-data range. RFCheck is the audit instantiation used in this paper, while repair and correction are used as interventions to test whether the detected failure can be reduced.

\subsection{Measurement-Consistency Score}

Let $m$ denote a sensing representation, and let $\mathcal{M}_{m,j}(x)$ be the $j$th measurement-side test for a sample $x$, where $j\in\{1,\ldots,J\}$. For each test, the audit calibrates an empirical threshold using held-out real samples from the same acquisition and preprocessing setting. Let
\[
\mathcal{S}_{\mathrm{real}}^{m,j} = 
\{\mathcal{M}_{m,j}(z):z\in\mathcal{D}_{\mathrm{cal}}^m\}
\]
denote the held-out real calibration scores for test $j$ of representation $m$.
In the implementation, $\widehat{Q}_{1-\alpha_j}$ denotes the order-statistic quantile with
$k_j=\lceil(n+1)(1-\alpha_j)\rceil$, where
$n=|\mathcal{S}_{\mathrm{real}}^{m,j}|$. In all reported settings, $k_j\le n$.
The calibrated threshold is
\begin{equation}
\label{eq:threshold}
\gamma_{m,j}^{(\alpha_j)} = 
\widehat{Q}_{1-\alpha_j}
\left(
\mathcal{S}_{\mathrm{real}}^{m,j}
\right).
\end{equation}
When all tests use the same branch level, we write $\alpha_j=\alpha$ for brevity.
The representation-level calibrated score is
\begin{equation}
\label{eq:generic_joint_score}
S_m(x) = 
\max_{j\in\{1,\ldots,J\}}
\frac{\mathcal{M}_{m,j}(x)}
{\gamma_{m,j}^{(\alpha_j)}} .
\end{equation}
A sample is flagged when $S_m(x)>1$. The maximum keeps a strong abnormality in one test from being averaged away by the others.

\subsubsection{Finite-Sample Calibration and Joint-Score Interpretation}
The threshold in Eq.~\eqref{eq:threshold} provides a finite-sample reference for each scalar measurement test. Let $Z_1,\ldots,Z_n$ be held-out real calibration samples from representation $m$ under the same acquisition and preprocessing setting, and let $Z_{n+1}$ be a future real sample from that setting. For a scalar test statistic $T(\cdot)$, let $\widehat{Q}_{1-\alpha}$ be the $k$th order statistic of $\{T(Z_i)\}_{i=1}^{n}$, with $k=\lceil(n+1)(1-\alpha)\rceil$ and $k\le n$.

\begin{theorem}[Finite-sample false-exceedance control]
\label{thm:false_exceedance}
If $T(Z_1),\ldots,T(Z_n),T(Z_{n+1})$ are exchangeable and ties are broken randomly or conservatively, then
\begin{equation}
\Pr\left(T(Z_{n+1})>\widehat{Q}_{1-\alpha}\right)
\le
\frac{n+1-k}{n+1}
\le
\alpha .
\end{equation}
\end{theorem}

The result follows from the uniform rank of $T(Z_{n+1})$ among the $n+1$
exchangeable scores; exceedance above the $k$th order statistic occurs with
probability at most $(n+1-k)/(n+1)\le\alpha$.

The finite-sample statement applies to each scalar measurement test. For the max score in Eq.~\eqref{eq:generic_joint_score}, the joint false-exceedance probability is controlled by a union bound:
\begin{equation}
\label{eq:joint_union_bound}
\Pr\left(S_m(Z_{n+1})>1\right)
\le
\sum_{j=1}^{J}\alpha_j .
\end{equation}
If all branches are calibrated at the same level $\alpha$, this bound becomes $J\alpha$. A target joint level $\alpha_{\mathrm{joint}}$ can be enforced by setting $\alpha_j=\alpha_{\mathrm{joint}}/J$. In this paper, we use the per-branch calibration as a conservative empirical measurement reference rather than as an exact $\alpha$-level joint test. The resulting thresholds are benchmark-specific finite-sample references, not universal measurement constants.

For CSI, the measured spectrum is a finite-bandwidth observation whose phase and support are shaped by the measurement chain \cite{kotaru2015spotfi}. A simplified observation model is
\begin{equation}
\label{eq:measurement_model}
H_{\mathrm{meas}}[k] = 
H(f_k)e^{-j2\pi f_k \tau_{\mathrm{off}}}e^{j\Phi_{\mathrm{rf}}[k]} + N[k].
\end{equation}
After occupied-tone embedding and inverse Fourier reconstruction, the CSI audit evaluates two complementary tests. The time-domain (TD) test measures precursor-like leakage before the aligned dominant arrival:
\begin{equation}
\label{eq:td_metric}
\mathcal{M}_{\mathrm{TD}} = 
\frac{\sum_{n \in\mathcal{N}_{\mathrm{pre}}} |\hat{h}[n]|^2}
{\sum_n |\hat{h}[n]|^2}.
\end{equation}
The frequency-domain (FD) test removes the dominant linear phase trend and combines residual phase irregularity with a local complex-continuation residual:
\begin{equation}
\label{eq:fd_metric}
\mathcal{M}_{\mathrm{FD}} = 
\tilde{\mathcal{M}}_{\mathrm{phase}}
+
\tilde{\mathcal{M}}_{\mathrm{local}} .
\end{equation}
The tilde denotes normalization by the held-out real calibration statistics, so the two FD components are combined on a comparable scale.
The CSI joint score is therefore
\begin{equation}
\label{eq:csi_score}
S_{\mathrm{CSI}} = 
\max\left(
\frac{\mathcal{M}_{\mathrm{TD}}}{\gamma_{\mathrm{TD}}^{(\alpha)}},
\frac{\mathcal{M}_{\mathrm{FD}}}{\gamma_{\mathrm{FD}}^{(\alpha)}}
\right).
\end{equation}
This score measures unusual finite-bandwidth CSI structure relative to the held-out real baseline. It is not a test of ideal continuous-time channel validity.

For FMCW gesture sensing, the same calibrated scoring principle is applied with tests matched to the radar representation. In the M-Gesture extension, each sample is a preprocessed radar tensor. The tests measure range leakage, range-Doppler (RD) local continuation, and receive-chain temporal residual. The corresponding score is
\begin{equation}
\label{eq:fmcw_score}
S_{\mathrm{FMCW}} = 
\max\left(
\frac{\mathcal{M}_{\mathrm{range}}}{\gamma_{\mathrm{range}}^{(\alpha)}},
\frac{\mathcal{M}_{\mathrm{RD}}}{\gamma_{\mathrm{RD}}^{(\alpha)}},
\frac{\mathcal{M}_{\mathrm{time}}}{\gamma_{\mathrm{time}}^{(\alpha)}}
\right).
\end{equation}
Thus, the shared component is the calibration workflow, while the measurement tests follow the sensing representation.

\subsection{Repair as a Failure-Reduction Reference}
The calibrated score first supports retention by measurement risk. Candidates are ranked by their calibrated measurement score under a fixed label-consistency and class-budget protocol. For the CSI anchor-residual setting, a candidate has the form
\begin{equation}
\hat{x}=x_{\mathrm{anchor}}+\Delta x,
\end{equation}
and a real-only classifier defines the training-side label-consistency indicator
\begin{equation}
\label{eq:label_pass}
P_{\mathrm{label}}(\hat{x},y)=
\mathbf{1}\left[\arg\max_c C_c(\hat{x})=y\right].
\end{equation}
This indicator is used only for candidate training samples, never for held-out test selection.

Retention can only choose among existing candidates. To test whether calibrated feedback is actionable, direct repair optimizes a small correction:
\begin{equation}
\min_{\delta}
\lambda_{\mathrm{meas}}\mathcal{L}_{\mathrm{meas}}(\hat{x}+\delta)
+
\lambda_{\mathrm{small}}\|\delta\|_2^2
+
\lambda_{\mathrm{lab}}\mathcal{L}_{\mathrm{lab}}(\hat{x}+\delta,y),
\end{equation}
where the measurement hinge penalty is
\begin{equation}
\mathcal{L}_{\mathrm{meas}}(\hat{x}+\delta) = 
\sum_j
\left[
\frac{\mathcal{M}_j(\hat{x}+\delta)}{\gamma_j}-1
\right]_+ .
\end{equation}
$\mathcal{L}_{\mathrm{lab}}$ is the classification loss from the fixed real-only classifier and is used only to preserve the candidate label during repair. Direct repair is used as a repair reference to test whether the detected measurement failure can be reduced while preserving task behavior. It is not presented as a deployable generator.

For dynamic FMCW gestures, repair should also avoid suppressing task-relevant motion. We therefore add a trajectory constraint when constructing the FMCW repair reference:
\begin{equation}
\mathcal{L}_{\mathrm{traj}} = 
\left\|
\psi_{\mathrm{traj}}(x+\delta)-\psi_{\mathrm{traj}}(x)
\right\|_2^2,
\end{equation}
where $\psi_{\mathrm{traj}}(\cdot)$ denotes range-Doppler motion descriptors. This term preserves gesture trajectories while still reducing the measurement score.

\begin{figure*}[htp!]
    \centering
    \includegraphics[width=0.95\linewidth]{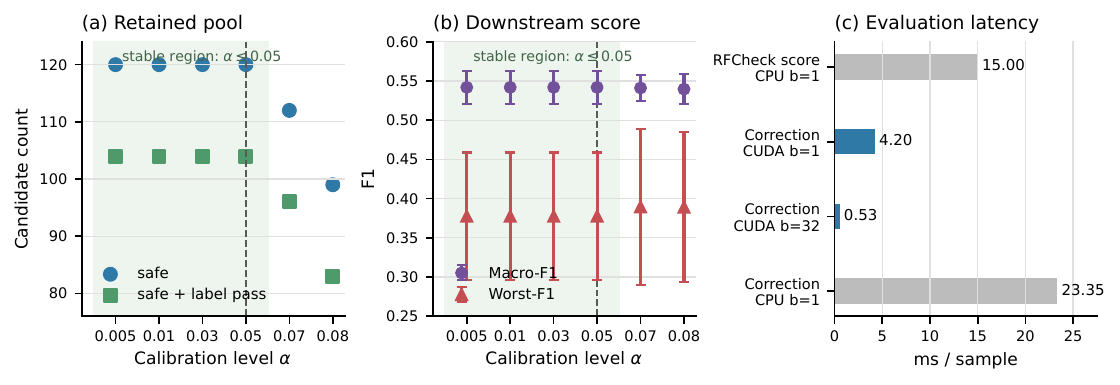}
    \caption{Calibration sensitivity and inference cost. Marker plots show that the selected pool and task score remain stable for $\alpha\le0.05$. The dashed line marks the default setting $\alpha=0.05$. Runtime is reported per sample.}
    \label{fig:alpha_runtime}
\end{figure*}

\subsection{Correcting Existing Synthetic Candidates}
The correction module starts from an existing synthetic proposal and uses a repair-reference target for supervision. It uses the proposal $x_{\mathrm{prop}}$ and the label $y$. It also uses normalized measurement scores $\mathbf{r}(x_{\mathrm{prop}})$ and measurement descriptors $\mathbf{m}(x_{\mathrm{prop}})$. The module predicts
\begin{equation}
\delta_{\theta} = 
T_{\theta}\!\left(
x_{\mathrm{prop}},y,\mathbf{r}(x_{\mathrm{prop}}),\mathbf{m}(x_{\mathrm{prop}})
\right),
\quad
x_{\mathrm{corr}}=x_{\mathrm{prop}}+\delta_\theta .
\end{equation}
The reference correction is
\begin{equation}
\delta^\star = x_{\mathrm{ref}} - x_{\mathrm{prop}},
\end{equation}
and the training objective is
\begin{equation}
\mathcal{L}_{\mathrm{corr}} = 
\underbrace{\|\delta_\theta-\delta^\star\|_1}_{\mathcal{L}_{\delta}}
+
\lambda_{\mathrm{tail}}\mathcal{L}_{\mathrm{tail}}
+
\lambda_{\mathrm{traj}}\mathcal{L}_{\mathrm{traj}}
+
\lambda_{\mathrm{small}}\|\delta_\theta\|_2^2 .
\end{equation}
For CSI correction, $\lambda_{\mathrm{traj}}=0$. The trajectory term is used only for FMCW samples to preserve range-Doppler motion structure.

The tail term focuses on near-threshold samples:
\begin{equation}
\mathcal{L}_{\mathrm{tail}} = 
\frac{1}{|\mathcal{B}_K|}
\sum_{i\in \mathcal{B}_K}
\sum_j
\left[
\frac{\mathcal{M}_j(x_{\mathrm{prop},i}+\delta_{\theta,i})}{\gamma_j}-\tau
\right]_+ ,
\end{equation}
where $\mathcal{B}_K$ is the highest-score subset within a minibatch, selected by the proposal-side normalized measurement score, and $\tau$ is a margin below the exceedance boundary.

This formulation separates two requirements for useful synthetic RF sensing samples. The synthetic candidate must preserve task-relevant sensing structure, while the calibrated measurement penalty constrains measurement validity. The module maps existing candidates toward repair-reference targets, so its role is limited to candidate pools that already contain usable sensing structure.

\subsection{Validation-Guided Candidate Selection}
Low measurement risk alone is insufficient for task performance. The final selection stage therefore combines calibrated measurement risk with task-relevant structure. For FMCW millimeter-wave, a hard filter first enforces low measurement score and motion consistency. The remaining samples are ranked on validation data using classifier margin and motion similarity:
\begin{equation}
\begin{aligned}
\mathrm{Rank}(x)
=&\ -w_1 S(x)
+w_2 \mathrm{Mar}(x)
-w_3 D_{\mathrm{traj}}(x) \\
&+w_4 \mathrm{Div}(x)
+w_5 E_{\mathrm{band}}(x)
-w_6 D_{\mathrm{anchor}}(x).
\end{aligned}
\end{equation}
Here $\mathrm{Mar}(x)$ is the validation classifier margin, $D_{\mathrm{traj}}(x)$ is the trajectory-distance penalty, $\mathrm{Div}(x)$ encourages diversity within the selected set, $E_{\mathrm{band}}(x)$ measures energy-band consistency, and $D_{\mathrm{anchor}}(x)$ measures distance to the anchor. All terms are normalized on the validation pool before weighting.

Weights and presets are selected on validation data only. The test set is never used for selection tuning. This stage reflects the main boundary of the paper. Calibrated measurement consistency is necessary for synthetic sensing samples, but task-relevant structure determines whether the sample is useful for training.

\section{Experimental Setup}
\label{sec:exp_setup}

The evaluation is organized around one central question. Can synthetic RF
sensing data fail measurement consistency even when they pass common validation
checks, and can this failure be reduced under matched measurement calibration?
Experiment~1 establishes the failure mode with controlled CSI and FMCW perturbations and tests whether selected summary statistics expose it.
Experiment~2 asks whether measurement risk still matters after label
consistency and class budget are fixed. Experiment~3 tests whether the detected
measurement failure can be reduced by a repair reference. Experiment~4 evaluates
whether proposal correction reduces measurement violations on
repeated and held-out synthetic candidates. Experiment~5 checks calibration sensitivity and runtime. It also tests whether the same calibration view extends to FMCW sensing.

\subsection{Benchmarks and Task Settings}
\label{subsec:setup_scope}

The primary evaluation focuses on CSI-based wireless sensing. Widar\cite{zhang2021widar3} is the main in-domain benchmark. Held-out real Widar CSI is used for controlled measurement-stress tests and perturbation construction. A low-resource Widar common3 protocol is then used for source audit and augmentation studies. ARIL\cite{wang2019joint} and WiMANS\cite{huang2024wimans} provide external checks under their own calibration and preprocessing. M-Gesture\cite{liu2021m} provides the FMCW millimeter-wave benchmark. It tests whether the same calibration principle can be instantiated for radar gesture data and whether validation-subject selection can recover useful low-risk subsets beyond CSI.

Within each benchmark, calibration data are disjoint from training and test data whenever task evaluation is performed. All RFCheck scores are computed under the corresponding benchmark-specific occupied-subcarrier pattern and preprocessing.

\subsection{Measurement Tests and Calibration}
\label{subsec:setup_scoring}

Each sample receives a TD score $\mathcal{M}_{\mathrm{TD}}$, an FD score $\mathcal{M}_{\mathrm{FD}}$, and a normalized joint score
\begin{equation}
S_{\mathrm{CSI}}=
\max\left(
\frac{\mathcal{M}_{\mathrm{TD}}}{\gamma_{\mathrm{TD}}},
\frac{\mathcal{M}_{\mathrm{FD}}}{\gamma_{\mathrm{FD}}}
\right),
\end{equation}
where $\gamma_{\mathrm{TD}}$ and $\gamma_{\mathrm{FD}}$ are empirical $(1-\alpha)$ quantiles from held-out real CSI with $\alpha=0.05$. For brevity, $\gamma_{\mathrm{TD}}$ and $\gamma_{\mathrm{FD}}$ denote $\gamma_{\mathrm{TD}}^{(0.05)}$ and $\gamma_{\mathrm{FD}}^{(0.05)}$ in the experiments. A sample is flagged when $S_{\mathrm{CSI}}>1$. Under exchangeability, the branch-wise thresholds admit the finite-sample interpretation in Section~\ref{sec:problem_framework}. In the present benchmark protocol, the joint threshold is interpreted conservatively as a benchmark-specific empirical reference point rather than as a universal guarantee.

The TD test is computed from delay-domain responses reconstructed from occupied-subcarrier CSI. The FD test is computed from local complex spectral continuation. For multiple receive links, scores are aggregated by worst-link maximization so that a clear inconsistency on one link is not hidden by averaging. Widar and WiMANS use the 30 Intel-style occupied Wi-Fi subcarriers embedded into a 64-tone grid. ARIL uses its benchmark-specific 52-subcarrier support. Sample scores use approximately 20 uniformly spaced frames and the 90th percentile over frames.

We also sweep $\alpha$ on the final CSI correction pool to check that the default
calibration level is not a fragile tuning choice. As shown in
Fig.~\ref{fig:alpha_runtime}, $\alpha\in[0.005,0.05]$ retains the same safe
candidate pool and gives the same macro-F1 under the
$20$-synthetic-samples-per-class protocol. Larger values begin to shrink the
retained pool, but the task score changes only within the observed seed
variation.

\begin{figure*}[htp!]
    \centering
    \includegraphics[width=0.92\textwidth]{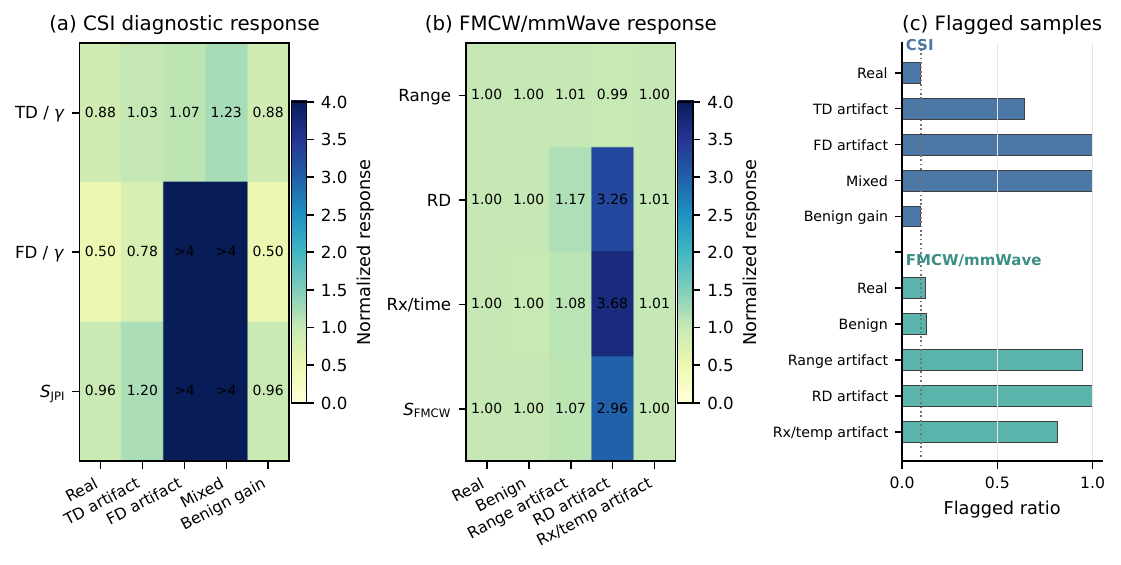}
    \caption{Controlled measurement-consistency tests across CSI and FMCW millimeter-wave. Heatmaps report normalized responses relative to the calibration reference, and bars report flagged ratios. The heatmaps show which test responds to each artifact, while the bar panel captures tail events whose mean response remains close to the reference.}
    \label{fig:cross_modality_diagnostic}
\end{figure*}

For FMCW, the same calibration protocol is retained while the test
definitions change. This experiment tests whether the idea transfers beyond CSI
rather than reusing CSI-specific tests. M-Gesture \texttt{long\_raw} samples are
scored by range leakage and range-Doppler continuity. A receive-chain temporal check is also used. The final flagged decision is made by $S_{\mathrm{FMCW}}>1$.
The exchangeable split is used to verify that held-out real and benign samples
remain close under the calibrated score, while subject-disjoint folds are used
as a harder domain-shift setting for augmentation.

\subsection{Downstream Augmentation Protocol}
\label{subsec:setup_subset_definition}

Experiments~3 and~4 use anchor-residual augmentation. A synthetic candidate is written as $\hat{x}=x_{\mathrm{anchor}}+\Delta x$, where $x_{\mathrm{anchor}}$ is a real CSI anchor and $\Delta x$ is a synthetic residual. Label consistency is checked by the real-only task classifier in Eq.~\eqref{eq:label_pass}. This check is applied only to the training-side candidate pool, never to held-out test samples.

The label-fixed study first restricts the candidate pool to label-consistency-passed
residuals and then compares random, low-score, high-score, low-FD, and high-FD
retention under the same class-balanced budget. It also audits three proposal families under the same ranking logic.
The weak generative adversarial network (GAN) source tests a low-quality boundary case, while the diffusion-style and refined sources test proposal pools with stronger sensing structure.

The repair study evaluates the score-guided repair reference as an offline
reference. The correction study evaluates repeated refined proposals and a
large held-out proposal pool. The Widar task protocol uses $120$ real training samples and retains $40$ synthetic samples per class unless otherwise stated. All task results are reported over five seeds using macro-F1 and worst-class F1 on held-out real test data. The WiMANS validation check uses five repeats of the official WiMANS protocol and reports exact-match accuracy, macro-F1, and
micro-F1.

For M-Gesture, we use motion-feature readouts built from range-Doppler and temporal descriptors. Repair with a trajectory constraint is compared with
measurement-only repair, and a selection rule uses hard
filtering and ranking to extract low-risk subsets from both the repair-reference pool and
the correction pool. This experiment is used to test whether RFCheck separates
the shared calibration component from the test design used for each representation.

For the selected-summary comparison, we compute the MMD using a radial basis function (RBF) kernel on magnitude and phase-difference summaries. The term ``matched'' therefore refers only to these selected summary descriptors and does not imply full distributional equivalence. The Smooth ramp and Local break pools are controlled perturbation pools and are not used for augmentation.

\subsection{Evaluation Protocol}
\label{subsec:setup_reporting}

The evaluation is organized by question. Controlled perturbations and
selected-summary comparisons test whether the failure exists. Label-fixed
retention tests whether it matters for augmentation. The repair reference and
proposal correction test whether it can be reduced. The final support study
checks calibration sensitivity, runtime, and the FMCW millimeter-wave extension.

All comparisons are interpreted within their own protocol and candidate pool.
Repeated runs report mean and standard deviation when available, while
single-summary rows are not used for seed-level statistical claims. For Widar,
the same real-only baseline is shared across retention, repair, and correction.
For M-Gesture, selection weights and budgets are fixed by validation subjects
only, and held-out test subjects are never used for selection tuning. We report
measurement metrics such as joint score and flagged ratio together with task
metrics such as macro-F1 and worst-class F1 because these quantities are not
interchangeable. A low RFCheck score indicates lower measurement risk, while
task performance must still be evaluated.

\begin{figure*}[htp!]
    \centering
    \includegraphics[width=0.96\linewidth]{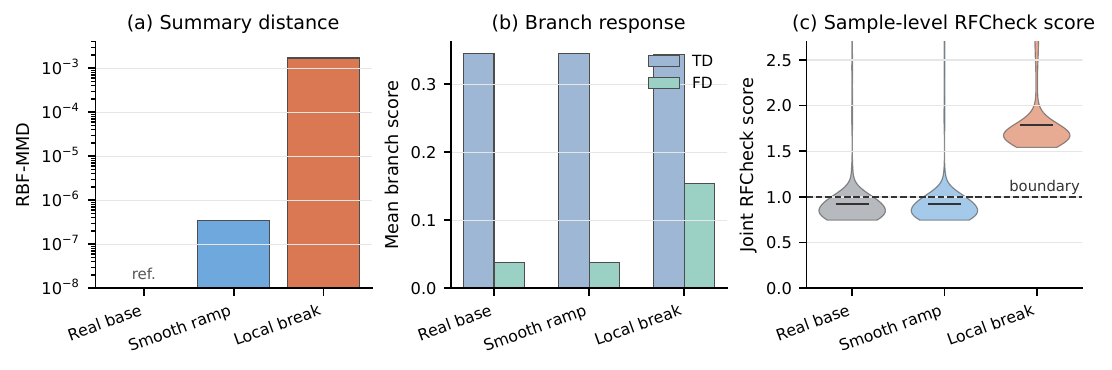}
    \caption{Selected-summary comparison and RFCheck measurement check on controlled CSI perturbations. Smooth ramp and Local break are constructed from the same real-base samples. The selected RBF-MMD summary reports aggregate shifts, while RFCheck identifies the measurement test responsible for structural inconsistency. Local break produces a clear FD and joint-score increase above the calibrated boundary. Dashed lines denote the calibrated boundary.}
    \label{fig:exp2_violin_fd}
\end{figure*}

Candidate pools are interpreted by role rather than by generator name. Weak
proposals test the boundary case where measurement correction cannot recover
missing task structure. Refined proposals test whether repair-reference targets
can correct samples that already retain task-relevant structure. The FMCW millimeter-wave proposal pool is handled similarly, but with range-Doppler and trajectory
structure in place of CSI residual structure.

We report computational cost on an Intel Xeon CPU server and an NVIDIA RTX 5090 GPU.
RFCheck scoring takes $15.00$ ms per CSI sample on CPU. Correction inference takes
$4.20$ ms per sample on GPU at batch size $1$ and $0.53$ ms per sample at batch
size $32$. Direct repair remains an offline optimization reference.

\section{Experimental Results}
\label{sec:exp_results}

\subsection{Evidence of Measurement-Consistency Failure}
\label{subsec:exp1_results}

Experiment~1 first tests whether measurement-consistency failure can be made visible under controlled RF perturbations. The goal is not only to validate a test statistic, but to show that synthetic-like samples can violate the real measurement structure in ways that require calibrated sample-level checking. We consider TD-canonical, FD-canonical, and mixed perturbations. The TD test is computed from delay-domain responses, the FD test is computed from local spectral continuation residuals, and the joint score is obtained by worst-link aggregation across receive chains.

Figure~\ref{fig:cross_modality_diagnostic} shows the main controlled response pattern. In CSI, TD scores increase mainly under TD-canonical perturbations, whereas FD scores increase mainly under FD-canonical and mixed perturbations. In FMCW, the heatmap and flagged-ratio panel should be read together. Range-Doppler artifacts produce a strong mean response, while range artifacts and receive-chain temporal artifacts are reflected more clearly by tail events. This confirms that the score captures measurement inconsistency rather than a single generic perturbation strength.

Per-RX analysis gives the same selectivity pattern. Under FD-canonical stress, the FD test increases consistently across receive chains, while benign perturbations cause only small score inflation in this setting. We keep the main text focused on the aggregate controlled response because it is the evidence used in the main argument.

The calibration/aggregation ablation explains why the final CSI score uses
empirical test quantiles and max aggregation. Mean aggregation dilutes
single-branch TD abnormalities. Under strong TD stress, quantile-mean flags only
$0.123$ of samples, whereas quantile-max flags $0.640$. The price of this
worst-branch sensitivity is that the clean joint flagged ratio is $0.093$ under
the two-branch max rule, above the nominal branch level. We therefore interpret
the joint score as a conservative union-style detector, not as an exact
$\alpha$-level joint test. The complete ablation is treated as supporting
evidence rather than as a separate main-text table.

Taken together, the controlled perturbation results show that the score captures domain-relevant structural deviations while remaining stable under benign controls. This test verifies selectivity under designed stresses. Coverage of all possible generator artifacts remains outside this controlled setting. The result is therefore used as evidence that measurement-consistency failure can be exposed before synthetic samples enter augmentation.

The selected-summary comparison tests whether aggregate summary distances are sufficient for measurement-consistency checking. The main comparison is conducted on held-out real Widar CSI. Three pools are constructed from the same real-base samples. Real base is the reference pool. Smooth ramp is a mild perturbation that remains close under the selected summary statistics. Local break introduces localized structural inconsistency and also increases the selected RBF-MMD summary. ARIL and WiMANS are used as external checks under their own calibration and preprocessing.

\begin{figure*}[htp!]
    \centering
    \includegraphics[width=0.96\linewidth]{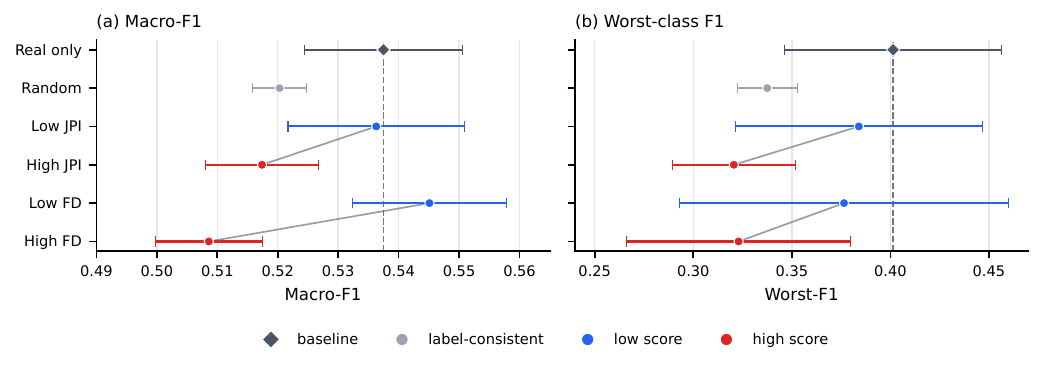}
    \caption{Label-fixed score-stratified retention on Widar. Error bars show seed standard deviation. With the same label-consistency acceptance and class-balanced budget, low-score retained residuals improve the low/high contrasts, especially for the weakest class.}
    \label{fig:exp3_score_stratified}
\end{figure*}

The main Widar result is shown in Fig.~\ref{fig:exp2_violin_fd}. The \emph{Smooth ramp} pool remains close to the real baseline under the selected summary distance. The \emph{Local break} pool shows an aggregate summary shift, but the summary distance does not identify the violated measurement structure or provide a sample-level decision. RFCheck supplies this missing information. Local break produces a clear FD increase and moves the joint score above the calibrated boundary, with FD mean increasing from about $0.038$ to $0.153$ and exceedance reaching $1.0000$. This shows that aggregate summary comparison is useful but insufficient for RF sensing measurement-consistency testing.

The same ordering is observed on ARIL and WiMANS under benchmark-specific calibration. Local break yields clear FD/joint separation, while Smooth ramp remains closer to the real baseline. The main text focuses on the Widar perturbation study because it is the dataset used for the repair and correction analysis.

The FMCW panel in Fig.~\ref{fig:cross_modality_diagnostic} provides the corresponding evidence beyond CSI. Under the exchangeable split, held-out real and weak-benign samples remain close, with flagged ratios near $0.12$, while range leakage and local range-Doppler breaks are strongly detected, with flagged ratios $0.9480$ and $1.0000$. This is the FMCW millimeter-wave counterpart to the CSI controlled test above. The same calibration logic transfers, but the test definitions change with the sensing representation.

The FMCW rows are intentionally reported under the exchangeable split
rather than only under subject-disjoint folds. Under subject-disjoint folds, real
samples can also exhibit elevated flagged rates because the calibration subjects
and test subjects differ in motion style and measurement statistics. We treat
that behavior as a measurement-domain shift stress test, not as a violation of
the finite-sample calibration argument. This distinction is important for using
RFCheck as a practical audit. Calibration should match the deployment
population whenever the goal is false-exceedance control, while subject-disjoint
stress tests reveal how sensitive the score is to sensing-condition changes.

\subsection{Measurement Risk Under Fixed Label and Budget}
\label{subsec:exp3_results}

Experiment~2 asks whether measurement inconsistency still matters after label
consistency has been fixed. This separates the proposed failure mode from
ordinary label errors. We first restrict the candidate pool to label-consistency-passed
anchor residuals and then compare low-score and high-score retained sets under
the same class-balanced budget. Performance is evaluated on held-out real test
data using macro-F1 and worst-class F1.

Fig.~\ref{fig:exp3_score_stratified}
shows the result. Random label-consistent retention does not recover the
real-only performance, especially for the weakest class. High-score retention
further degrades the weakest class. Label-high-score reaches only
$0.3206\pm0.0311$ worst-class F1 and label-high-FD reaches
$0.3230\pm0.0569$. In contrast, label-low-score raises worst-class F1 relative
to high-score retention to $0.3840\pm0.0628$, and label-low-FD gives the
highest macro-F1, $0.5451\pm0.0128$. Because all synthetic rows satisfy the
same label-consistency acceptance condition, this gap isolates measurement consistency as an additional variable beyond label-consistency filtering.

The low/high contrasts are the important evidence. Low-score retention improves
worst-class F1 over high-score retention by $0.0634$, while low-FD retention
improves macro-F1 over high-FD retention by $0.0365$. A conservative seed-level
confidence check gives a clear interval for the low-FD macro-F1 contrast
($95\%$ CI roughly $[0.020,0.053]$ under the repeated-seed summaries), whereas
the low-score weakest-class contrast is directionally positive but has wider seed
uncertainty. RFCheck should therefore not be interpreted as a task-performance
predictor by itself, but it does provide a measurement-side ordering that is not
available from the label filter itself. A compact validation check with the official WiMANS backbone reaches the same conclusion. RFCheck-only selection
removes retained flagged samples and obtains the highest micro-F1 among the
tested selection variants, while RFCheck combined with the label-consistency
filter matches the real-only macro-F1 and reduces the flagged ratio to $0.0111$.

\begin{figure*}[thp!]
    \centering
    \includegraphics[width=0.95\linewidth]{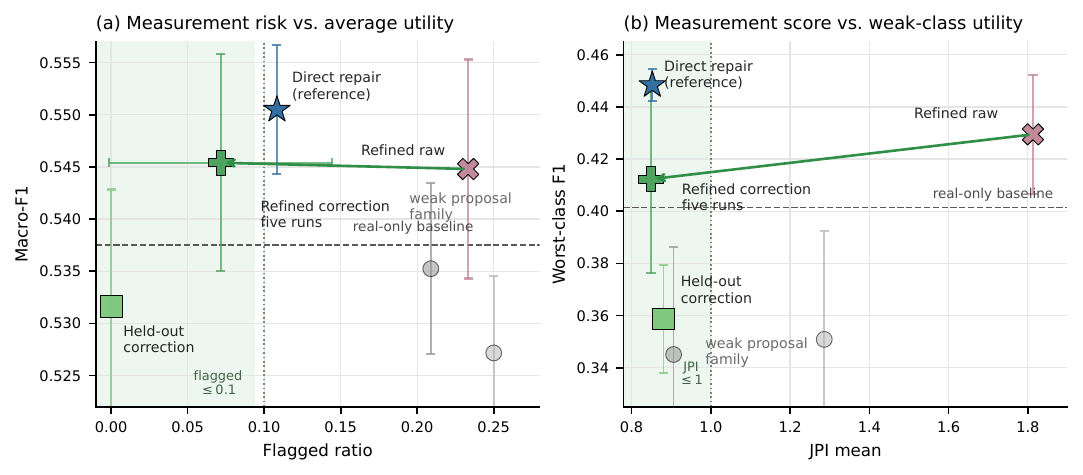}
    \caption{Measurement risk and task-performance tradeoff of proposal correction on Widar. Correction reduces measurement risk for refined candidates, while the weak-class result shows that measurement-side safety and uniform class recovery remain distinct objectives. Direct repair is shown only as an offline reference.}
    \label{fig:exp4_repair_summary}
\end{figure*}

The same source-audit experiment extends the label-fixed analysis to several
synthetic sources. For refined candidates at the 20/class budget, low-score
selection reduces joint score from $2.415$ to $1.211$ and flagged ratio from $0.333$ to
$0.133$, with macro-F1 and worst-class F1 changing from $0.5280/0.3371$ to
$0.5379/0.3711$. For the diffusion-style source, low-score selection lowers joint score
from $0.978$ to $0.826$ and removes flagged samples at 20/class, while the
task gap is small. GAN low/high subsets have low calibrated risk but weak
and similar task performance. We therefore summarize these rows in text rather than
as a separate table. Their main message is qualitative. RFCheck separates
measurement risk across sources, but task performance still depends on
proposal task structure and budget.

\begin{table*}[hbp!]
\centering
\caption{Main CSI evidence for measurement-consistency failure and mitigation on Widar. Audit rows report the refined-source 20/class low/high RFCheck contrast. Repair rows use the label-passed residual pool. The held-out correction row contains 120 samples with a 40/40/40 class histogram, no unsafe or structure fallback, and maximum joint score below one in this held-out selection.}
\label{tab:main_csi}
\begin{tblr}{
  width = \linewidth,
  colspec = {Q[65]Q[170]Q[105]Q[105]Q[105]Q[127]Q[127]},
  cells = {c},
  row{1} = {font=\bfseries},
  cell{3}{1} = {r=2}{},
  cell{5}{1} = {r=2}{},
  cell{7}{1} = {r=2}{},
  hlines,
  vlines,
}
Block    & Method                           & Joint score              & Flagged           & Label pass        & Macro-F1                   & Worst-F1                   \\
Baseline & real-only                        & –                 & –                 & –                 & $0.5375\pm0.0130$          & $0.4014\pm0.0550$          \\
Audit    & low-score retained                 & $1.2106$          & $0.1333$          & $1.0000$          & $0.5379\pm0.0207$          & $0.3711\pm0.0873$          \\
         & high-score retained                & $2.4150$          & $0.3333$          & $1.0000$          & $0.5280\pm0.0213$          & $0.3371\pm0.1184$          \\
Repair   & raw label-passed residuals       & $2.4666$          & $0.6833$          & $1.0000$          & $0.5203$                   & $0.3375$                   \\
         & full direct repair               & $0.8523$          & $0.1083$          & –                 & $0.5505\pm0.0062$          & $0.4484\pm0.0062$          \\
{Correction\\~} & refined proposal correction, five seeds & $0.8489\pm0.0115$ & $0.0717\pm0.0728$ & $0.8800\pm0.0347$ & $0.5454\pm0.0104$          & $0.4123\pm0.0361$          \\
         & held-out correction selected            & $0.8802$          & $0.0000$          & $1.0000$          & $0.5316\pm0.0112$          & $0.3587\pm0.0206$          
\end{tblr}
\end{table*}

\subsection{Reducing Measurement Inconsistency With a Repair Reference}
\label{subsec:repair_results}

\begin{figure*}[htp!]
    \centering
    \includegraphics[width=0.85\linewidth]{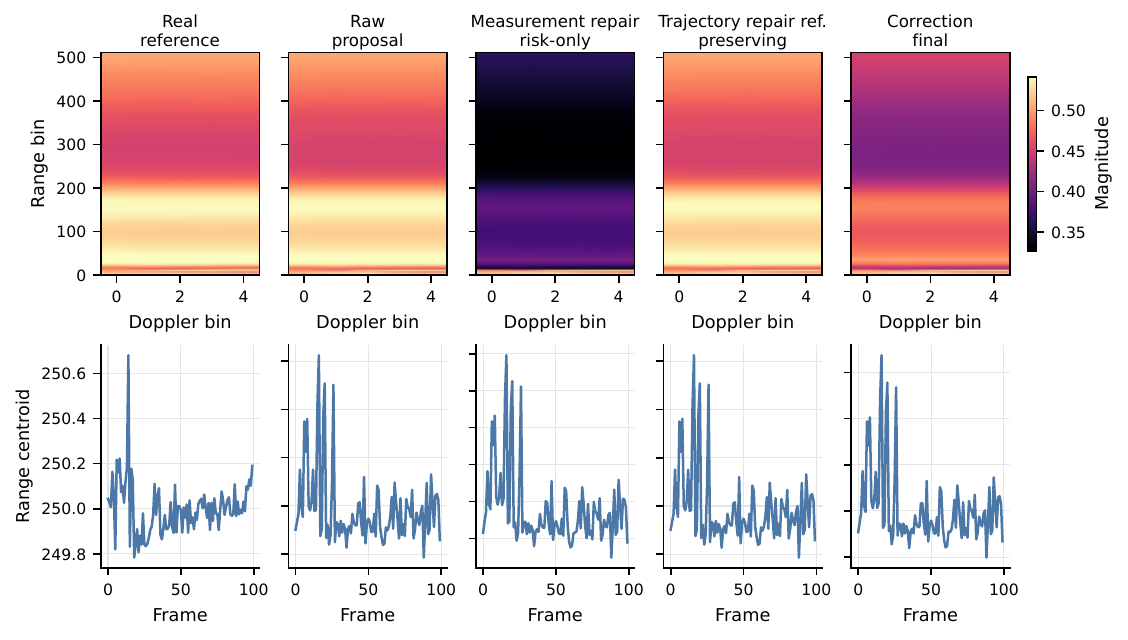}
    \caption{FMCW millimeter-wave qualitative repair and correction visualization. Range-Doppler magnitude maps and range-centroid trajectories illustrate how measurement-only repair, repair with a trajectory constraint, and the final correction change the proposal under the M-Gesture representation.}
    \label{fig:fmcw_qualitative_repair_correction}
\end{figure*}

Experiment~3 tests whether the detected failure can be reduced. A measurement audit would be incomplete if it only exposed risky samples without showing whether the violation can be mitigated. We therefore use direct repair as an offline intervention to test whether reducing measurement violations can preserve task behavior. In CSI, the raw
label-passed residual pool has joint score $2.4666$, flagged ratio $0.6833$, macro-F1
$0.5203$, and worst-class F1 $0.3375$, showing that label pass alone is not
enough. A task-only restoration baseline lowers joint score to $1.5725$ but still leaves a flagged ratio of $0.2250$. Full direct repair is the key reference. As summarized
in Table~\ref{tab:main_csi}, it reduces joint score to $0.8523$, lowers flagged ratio to
$0.1083$, and improves the mean macro-F1 and worst-class F1 to $0.5505/0.4484$. A
conservative seed-level contrast against real-only is positive in mean but has
wide uncertainty. The macro-F1 gain is $+0.0130$ with CI roughly $[-0.023,0.049]$.
The worst-class F1 gain is $+0.0470$ with CI roughly $[-0.090,0.184]$. This row is used as
upper/reference evidence for actionability rather than as a formal claim of
task superiority. This result shows that the detected failure is reducible under a repair intervention, rather than merely observable by an audit. In
FMCW, the corresponding repair chain is shown visually in
Fig.~\ref{fig:fmcw_qualitative_repair_correction}, measurement-only repair lowers
flagged rate, but repair with the trajectory constraint is the cleaner reference because
it constrains motion structure while keeping the low-risk behavior.

\subsection{Correction on Held-Out Synthetic Proposals}
\label{subsec:exp4_repair}

Experiment~4 moves from the repair reference to correction of existing proposals.
This experiment tests the main boundary of the paper. Measurement inconsistency
can be reduced when the proposal pool already contains task-relevant sensing
structure, but correction cannot recover arbitrary weak proposals. We compare a direct-repair reference, weak GAN proposal rows that expose the boundary of proposal correction, and a refined proposal row on which correction is evaluated again.
Table~\ref{tab:main_csi} and Fig.~\ref{fig:exp4_repair_summary} summarize the
main CSI correction evidence. Direct repair is the offline reference. It reduces
the flagged ratio to $0.1083$ and raises macro-F1 and worst-class F1 above
real-only in mean, although the seed-level interval is wide. Weak GAN proposals remain limited after score-guided correction. Their measurement risk decreases,
but the task readout does not approach the repair reference. The refined
proposal family separates proposal quality from RFCheck-guided correction. Raw
refined proposals have useful task structure, with macro-F1
$0.5448\pm0.0105$ and worst-class F1 $0.4295\pm0.0228$, but remain
measurement-risky, with mean joint score $1.8128$ and flagged ratio $0.2333$.

To avoid relying on a single positive run, we trained five correction runs under
the same split, proposal, repair
target, and task protocol. Across five runs, refined correction consistently lowers measurement risk. 
The mean joint score is $0.8489\pm0.0115$, and the flagged ratio is $0.0717\pm0.0728$. 
The corresponding macro-F1 and worst-class F1 are $0.5454\pm0.0104$ and $0.4123\pm0.0361$. At the
same time, the nonzero flagged-rate standard deviation shows that threshold-tail
control remains correction-run sensitive. The best confirmed run is not used as a
run-level superiority claim.

The correction ablation is reported as a compact trend in the correction summary figure
because it is a tradeoff result rather than a leaderboard. Removing the RFCheck
energy term yields high task F1 in this split, but it raises joint score from
$0.8489$ to $1.8386$ and the flagged ratio from $0.0717$ to $0.2250$. Thus
utility-oriented objectives alone do not control calibrated measurement tails.
The other ablations change the measurement risk and task-performance tradeoff rather than monotonically
improving every metric, which is consistent with the central claim that
measurement consistency and task performance are complementary axes.

We further evaluate correction on a large held-out proposal pool that is not used
to fit the correction module. The held-out pool contains $1140$ candidates, with
$26.32\%$ flagged before selection. Under the fixed $40$-per-class budget,
RFCheck selects $120$ candidates with a balanced $40/40/40$ class histogram. 
No selected sample is flagged, and the mean joint score is $0.8802$. 
All selected samples also pass the label check. This selected set largely preserves average utility,
with macro-F1 $0.5316\pm0.0112$ relative to the real-only
$0.5375\pm0.0130$. The weakest class remains limited. Class 0 reaches
$0.3587\pm0.0206$, while a lightweight class-weighted readout raises it to
$0.4029\pm0.0158$ without changing RFCheck or the correction step. Thus the
held-out row supports correction that reduces measurement violations,
not uniform class-wise recovery or guaranteed task improvement.

On the same refined-correction candidate pool, RFCheck ranking gives the
lowest actual joint score after selection ($0.7799$). A task-confidence proxy gives the
strongest task F1 but with higher measurement risk, while RBF-MMD summary
ranking and generic out-of-distribution scoring give weaker measurement risk and task-performance tradeoffs. These
selection baselines support the intended positioning. Measurement-calibrated
scoring is complementary to task confidence rather than a replacement for it.

\begin{table*}[htp!]

\centering
\caption{FMCW millimeter-wave summary on M-Gesture \texttt{long\_raw}. Rows use
validation-subject selection. The high fallback and trajectory gap of correction
selection indicate that low measurement risk alone is insufficient for dynamic
RF augmentation.}
\label{tab:exp5_fmcw_summary}
\begin{tblr}{
  width = \linewidth,
  colspec = {Q[270]Q[79]Q[107]Q[167]Q[207]Q[81]},
  cells = {c},
  row{1} = {font=\bfseries},
  hlines,
  vlines,
}
Method                             & Flagged & Trajectory gap & Macro-F1          & Worst-F1          & Fallback \\
real-only, SVM                     & –       & –         & $0.2376\pm0.0310$ & $0.1115\pm0.0919$ & –        \\
raw proposal                       & 0.5738  & –         & –                 & –                 & –        \\
trajectory repair ref.             & 0.0100  & 0.1722    & –                 & –                 & –        \\
trajectory repair + selection, budget 10 & 0.0000  & 0.0150    & $0.2627\pm0.0282$ & $0.1357\pm0.0525$ & 0.0050   \\
correction selection, budget 40              & 0.0100  & 0.9475    & $0.2516\pm0.0356$ & $0.1777\pm0.0522$ & 0.7238   
\end{tblr}
\end{table*}

\subsection{Calibration Sensitivity and Cross-Representation Evidence}
\label{subsec:exp5_fmcw}

Experiment~5 checks whether the finding is tied to CSI or reflects a broader RF
measurement issue. We keep the calibration principle fixed and change the
measurement tests for FMCW gesture sensing.

The sensitivity and latency measurements in Fig.~\ref{fig:alpha_runtime}
support the practical use of RFCheck as an audit and correction layer. On the
final CSI correction pool, $\alpha\in[0.005,0.05]$ retains the same safe candidate set
and gives the same macro-F1 under the 20/class protocol. Looser
thresholds begin to shrink and unbalance the retained set. RFCheck scoring is a
deterministic measurement pass, while correction inference is substantially cheaper than
direct repair and is therefore the path used for proposal correction at test
time.

We next ask whether the same calibration principle can be instantiated on a
heterogeneous RF sensing representation. The cross-representation benchmark is M-Gesture
\texttt{long\_raw}, where each sample is a preprocessed FMCW millimeter-wave tensor with
dominant task structure in range-Doppler motion and receive-chain temporal
organization. The role of this experiment is to separate the shared calibration
component of RFCheck from the test design used for each representation.

Under exchangeable calibration, held-out real and weak-benign FMCW samples remain
near the calibrated range, while range leakage and range-Doppler breaks are
strongly detected. Fig.~\ref{fig:fmcw_qualitative_repair_correction} shows why
task preservation matters. Measurement-only repair lowers flagged ratio but
distorts motion structure, whereas repair with the trajectory constraint keeps low-risk
behavior while preserving the trajectory more cleanly. Correction approaches the
low-flagged behavior of the repair reference but remains less complete, as shown
by the larger trajectory gap and fallback ratio in
Table~\ref{tab:exp5_fmcw_summary}.

The task-fixed FMCW ranking mirrors the CSI low/high comparison. Under the SVM readout with budget 10, low-score correction reduces flagged ratio from $0.0400$ to $0.0000$ and improves macro-F1 and worst-class F1 from $0.2263/0.1236$ to $0.2820/0.1918$. The RF readout shows the same macro-F1 trend, although worst-class F1 remains mixed. Table~\ref{tab:exp5_fmcw_summary} therefore supports the same measurement-consistency argument under a protocol with validation-subject selection rather than broad augmentation across all readouts or
budgets. The large fallback ratio ($0.7238$) and trajectory gap ($0.9475$) in the correction-selection row mark the main boundary of the FMCW study. Correction can reduce measurement-side violations, as reflected by the low flagged ratio, but it cannot recover coherent range-Doppler motion when the original synthetic candidate pool does not retain sufficient task-relevant sensing structure. Low measurement risk is therefore necessary but not sufficient for useful dynamic RF
augmentation.

\section{Discussion, Limitations, and Conclusion}
\label{sec:discussion_conclusion}

The controlled measurement tests provide the strongest evidence that synthetic-like RF samples can violate representation-specific measurement structure while remaining difficult to identify from selected summaries or label checks. RFCheck provides a calibrated audit for making this failure observable at the sample level. The repair experiments then test whether the detected failure can be reduced under an offline intervention. Correction applies the repair reference to proposals whose sensing structure is already useful for the task. The FMCW millimeter-wave experiment tests whether the same failure-and-calibration view can be instantiated beyond CSI.

Measurement consistency is complementary to label correctness and task performance. In the Widar retention study, all retained candidates pass the same label-consistency check, but low-score candidates outperform high-score candidates. The correction results show the same boundary. Candidate pools with limited task-relevant sensing structure remain difficult to correct, whereas refined candidates that retain useful sensing structure allow correction to reduce measurement violations while preserving average performance. In the larger held-out study, correction produces a class-balanced and label-consistent selected set with no selected sample above the RFCheck threshold, but class 0 remains the bottleneck under the default classifier. The correction claim is therefore proposal-based and centered on reducing measurement violations rather than on universal task improvement. The FMCW millimeter-wave results further show that RFCheck separates the shared calibration component from the test design used for each representation.

Several limitations follow. Empirical thresholds depend on the calibration data, hardware state, and preprocessing.
$S_{\mathrm{CSI}}=1$ and $S_{\mathrm{FMCW}}=1$ are benchmark-specific references, not universal measurement boundaries. RFCheck measures measurement consistency, not label correctness or task accuracy. The correction step is limited when the synthetic candidate pool lacks task-relevant sensing structure. The FMCW millimeter-wave results support cross-representation RF sensing evidence under a protocol with validation-subject selection, not broad RF generation across other RF modalities.

This paper showed that synthetic RF sensing data can fail measurement consistency under matched measurement settings. CSI provides the primary evidence through TD precursor leakage and FD local continuation. Controlled experiments, selected-summary comparisons, and label-fixed retention show that this failure can be missed by common checks. The score-guided repair reference reduces the flagged ratio to $10.83\%$ while preserving mean task performance. Proposal correction then produces low-risk, class-balanced held-out candidates while largely preserving average performance. Overall, the reliability of synthetic RF sensing data for augmentation depends not only on task performance, but also on measurement consistency with held-out real samples collected and processed under the same sensing pipeline.

{
\small
\bibliographystyle{IEEEtran}
\bibliography{bibtex/bib/IEEEabrv,bibtex/bib/IEEEreference}

@ARTICLE{9737357,
  author={Liu, Fan and Cui, Yuanhao and Masouros, Christos and Xu, Jie and Han, Tony Xiao and Eldar, Yonina C. and Buzzi, Stefano},
  journal={{IEEE} J. Sel. Areas Commun.}, 
  title={Integrated Sensing and Communications: Toward Dual-Functional Wireless Networks for 6G and Beyond}, 
  year={2022},
  volume={40},
  number={6},
  pages={1728-1767},
  doi={10.1109/JSAC.2022.3156632}}

@inproceedings{hayashi2021radarnet,
  title={RadarNet: Efficient Gesture Recognition Technique Utilizing a Miniature Radar Sensor},
  author={Hayashi, Eri and Lien, Jamie and Gillian, Nicholas and Giusti, Lorenzo and Weber, Daniel and Yamanaka, Jun and Bedal, Luca and Poupyrev, Ivan},
  booktitle={Proceedings of the 2021 CHI Conference on Human Factors in Computing Systems},
  pages={1--14},
  year={2021}
}

@article{grobelny2022mm,
  title={MM-Wave radar-based recognition of multiple hand gestures using long short-term memory (LSTM) neural network},
  author={Grobelny, Piotr and Narbudowicz, Adam},
  journal={Electronics},
  volume={11},
  number={5},
  pages={787},
  year={2022},
  publisher={MDPI}
}

@article{liu2021m,
  title={M-gesture: Person-independent real-time in-air gesture recognition using commodity millimeter wave radar},
  author={Liu, Haipeng and Zhou, Anfu and Dong, Zihe and Sun, Yuyang and Zhang, Jiahe and Liu, Liang and Ma, Huadong and Liu, Jianhua and Yang, Ning},
  journal={IEEE Internet Things J.},
  volume={9},
  number={5},
  pages={3397--3415},
  year={2021},
  publisher={IEEE}
}

@INPROCEEDINGS{10437154,
  author={Sengupta, Ushnish and Jao, Chinkuo and Bernacchia, Alberto and Vakili, Sattar and Shiu, Da-shan},
  booktitle={GLOBECOM 2023 - 2023 IEEE Global Communications Conference}, 
  title={Generative Diffusion Models for Radio Wireless Channel Modelling and Sampling}, 
  year={2023},
  volume={},
  number={},
  pages={4779-4784},
  doi={10.1109/GLOBECOM54140.2023.10437154}}

@article{wen2025generative,
  title={Generative {AI} for Data Augmentation in Wireless Networks: Analysis, Applications, and Case Study},
  author={Wen, Jinbo and Kang, Jiawen and Niyato, Dusit and Zhang, Yang and Wang, Jiacheng and Sikdar, Biplab and Zhang, Ping},
  journal={IEEE Wireless Communications},
  pages={1--10},
  year={2025},
  doi={10.1109/MWC.2025.3610574},
  publisher={IEEE}
}

@article{chen2023cross,
  title={Cross-Domain {WiFi} Sensing With Channel State Information: A Survey},
  author={Chen, Chen and Zhou, Gang and Lin, Youfang},
  journal={ACM Computing Surveys},
  volume={55},
  number={11},
  pages={1--37},
  year={2023},
  doi={10.1145/3570325},
  publisher={ACM New York, NY}
}

@inproceedings{gong2025data,
  title={Data Can Speak for Itself: Quality-guided Utilization of Wireless Synthetic Data},
  author={Gong, Chen and Liang, Bo and Gao, Wei and Xu, Chenren},
  booktitle={Proceedings of the 23rd Annual International Conference on Mobile Systems, Applications and Services},
  pages={209--222},
  year={2025},
  doi={10.1145/3711875.3729123}
}

@article{bock2025physics,
  title={Physics-informed generative modeling of wireless channels},
  author={B{\"o}ck, Benedikt and Oeldemann, Andreas and Mayer, Timo and Rossetto, Francesco and Utschick, Wolfgang},
  journal={Proceedings of Machine Learning Research},
  volume={267},
  pages={4602--4626},
  year={2025}
}

@article{zhang2026integrated,
  title={Integrated Sensing and Communications Over the Years: An Evolution Perspective},
  author={Zhang, Di and Cui, Yuanhao and Cao, Xiaowen and Su, Nanchi and Gong, Yi and Liu, Fan and Yuan, Weijie and Jing, Xiaojun and Zhang, J. Andrew and Xu, Jie and Masouros, Christos and Niyato, Dusit and Di Renzo, Marco},
  journal={IEEE Communications Surveys \& Tutorials},
  year={2026},
  pages={1--1},
  doi={10.1109/COMST.2026.3655674},
  publisher={IEEE}
}

@article{alkhateeb2019deepmimo,
  title={{DeepMIMO}: A Generic Deep Learning Dataset for Millimeter Wave and Massive {MIMO} Applications},
  author={Alkhateeb, Ahmed},
  journal={arXiv preprint arXiv:1902.06435},
  year={2019},
  doi={10.48550/arXiv.1902.06435}
}

@inproceedings{wang2015understanding,
  title={Understanding and modeling of wifi signal based human activity recognition},
  author={Wang, Wei and Liu, Alex X and Shahzad, Muhammad and Ling, Kang and Lu, Sanglu},
  booktitle={Proceedings of the 21st annual international conference on mobile computing and networking},
  pages={65--76},
  year={2015}
}

@article{van2024generative,
  title={Generative AI for physical layer communications: A survey},
  author={Van Huynh, Nguyen and Wang, Jiacheng and Du, Hongyang and Hoang, Dinh Thai and Niyato, Dusit and Nguyen, Diep N and Kim, Dong In and Letaief, Khaled B},
  journal={IEEE Transactions on Cognitive Communications and Networking},
  volume={10},
  number={3},
  pages={706--728},
  year={2024},
  publisher={IEEE}
}

@article{gretton2012kernel,
  title={A kernel two-sample test},
  author={Gretton, Arthur and Borgwardt, Karsten M and Rasch, Malte J and Sch{\"o}lkopf, Bernhard and Smola, Alexander},
  journal={The journal of machine learning research},
  volume={13},
  number={1},
  pages={723--773},
  year={2012},
  publisher={JMLR. org}
}

@ARTICLE{lan2024bullydetect,
  author={Lan, Bo and Wang, Fei and Xia, Lekun and Nai, Fan and Nie, Shiqiang and Ding, Han and Han, Jinsong},
  journal={IEEE Internet of Things Journal}, 
  title={BullyDetect: Detecting School Physical Bullying With Wi-Fi and Deep Wavelet Transformer}, 
  year={2024},
  doi={10.1109/JIOT.2024.3486071}
}

@inproceedings{chi2024rf,
  title={{RF-Diffusion}: Radio Signal Generation via Time-Frequency Diffusion},
  author={Chi, Guoxuan and Yang, Zheng and Wu, Chenshu and Xu, Jingao and Gao, Yuchong and Liu, Yunhao and Han, Tony Xiao},
  booktitle={Proceedings of the 30th Annual International Conference on Mobile Computing and Networking},
  pages={77--92},
  year={2024}
}

@inproceedings{huang2024wimans,
  title={{WiMANS}: A Benchmark Dataset for {WiFi}-Based Multi-User Activity Sensing},
  author={Huang, Shuokang and Li, Kaihan and You, Di and Chen, Yichong and Lin, Arvin and Liu, Siying and Li, Xiaohui and McCann, Julie A},
  booktitle={Computer Vision -- ECCV 2024},
  pages={72--91},
  year={2024},
  doi={10.1007/978-3-031-72946-1_5},
  organization={Springer}
}

@article{wang2024xrf55,
  title={{XRF55}: A Radio Frequency Dataset for Human Indoor Action Analysis},
  author={Wang, Fei and Lv, Yizhe and Zhu, Mengdie and Ding, Han and Han, Jinsong},
  journal={Proceedings of the ACM on Interactive, Mobile, Wearable and Ubiquitous Technologies},
  volume={8},
  number={1},
  pages={1--34},
  year={2024},
  doi={10.1145/3643543},
  publisher={ACM New York, NY, USA}
}

@article{tan2022commodity,
  title={Commodity {WiFi} Sensing in Ten Years: Status, Challenges, and Opportunities},
  author={Tan, Sheng and Ren, Yili and Yang, Jie and Chen, Yingying},
  journal={IEEE Internet of Things Journal},
  volume={9},
  number={18},
  pages={17832--17843},
  year={2022},
  publisher={IEEE}
}

@article{wen2024survey,
  title={A survey on integrated sensing, communication, and computation},
  author={Wen, Dingzhu and Zhou, Yong and Li, Xiaoyang and Shi, Yuanming and Huang, Kaibin and Letaief, Khaled B},
  journal={IEEE Communications Surveys \& Tutorials},
  year={2024},
  publisher={IEEE}
}

@article{zhu2025csi,
  title={{CSI}-Bench: A Large-Scale In-the-Wild Dataset for Multi-Task {WiFi} Sensing},
  author={Zhu, Guozhen and Hu, Yuqian and Gao, Weihang and Wang, Wei-Hsiang and Wang, Beibei and Liu, KJ},
  journal={arXiv preprint arXiv:2505.21866},
  year={2025}
}

@article{letafati2023diffusion,
  title={Diffusion models for wireless communications},
  author={Letafati, Mehdi and Ali, Samad and Latva-aho, Matti},
  journal={arXiv preprint arXiv:2310.07312},
  year={2023}
}

@article{hou2024rfboost,
  title={{RFBoost}: Understanding and Boosting Deep {WiFi} Sensing via Physical Data Augmentation},
  author={Hou, Weiying and Wu, Chenshu},
  journal={Proceedings of the ACM on Interactive, Mobile, Wearable and Ubiquitous Technologies},
  volume={8},
  number={2},
  pages={1--26},
  year={2024},
  doi={10.1145/3659620},
  publisher={ACM New York, NY, USA}
}

@article{zhang2021widar3,
  title={{Widar3.0}: Zero-Effort Cross-Domain Gesture Recognition With {Wi-Fi}},
  author={Zhang, Yi and Zheng, Yue and Qian, Kun and Zhang, Guidong and Liu, Yunhao and Wu, Chenshu and Yang, Zheng},
  journal={IEEE Transactions on Pattern Analysis and Machine Intelligence},
  volume={44},
  number={11},
  pages={8671--8688},
  year={2022},
  doi={10.1109/TPAMI.2021.3105387},
  publisher={IEEE}
}

@article{wang2019joint,
  title={Joint Activity Recognition and Indoor Localization With {WiFi} Fingerprints},
  author={Wang, Fei and Feng, Jianwei and Zhao, Yinliang and Zhang, Xiaobin and Zhang, Shiyuan and Han, Jinsong},
  journal={IEEE Access},
  volume={7},
  pages={80058--80068},
  year={2019},
  doi={10.1109/ACCESS.2019.2923743},
  publisher={IEEE}
}

@inproceedings{kotaru2015spotfi,
  title={Spotfi: Decimeter level localization using wifi},
  author={Kotaru, Manikanta and Joshi, Kiran and Bharadia, Dinesh and Katti, Sachin},
  booktitle={Proceedings of the 2015 ACM conference on special interest group on data communication},
  pages={269--282},
  year={2015}
}
}

\end{document}